\documentstyle[12pt,epsf]{article}
\makeatletter
\@addtoreset{equation}{section}

\makeatother

\begin{document}

\begin{flushright}

Oct 2001

KEK-TH-786

\end{flushright}

\begin{center}

\vspace{5cm}

{\Large Properties of String Theory}

\vspace{1cm}

{\Large on Kaluza-Klein Melvin Background} 

\vspace{2cm}

Takao Suyama \footnote{e-mail address : tsuyama@post.kek.jp}

\vspace{1cm}

{\it Theory Group, KEK}

{\it Tsukuba, Ibaraki 305-0801, Japan}

\vspace{4cm}

{\bf Abstract} 

\end{center}

We investigate some properties of string theory on Kaluza-Klein Melvin background. 
We discuss what happens when closed string tachyons in the theory condense, by using a 
linear sigma model of the Kaluza-Klein Melvin background. 

\newpage

\vspace{1cm}

\section{Introduction}

\vspace{5mm}

The stability of vacua of string theory is one of the main topics in recent researches. 
It is well-known that there are infinitely many perturbative vacua in string theory, and 
some of them contain tachyons in their spectra. 
The existence of the tachyons has been considered as an indication of an instability of the 
vacuum, not of the inconsistency. 
However, such vacua have not been discussed, although some attempts were made \cite{TC}, 
because of the lack of the knowledge of the mechanism for the stabilization of them. 

Now we know the fate of perturbative vacua which contain tachyons in the open string 
sector \cite{open}. 
The existence of open string tachyons is related to the existence of an unstable D-brane 
system. 
The tachyons acquire nonzero vevs, at which the effective potential takes a minimum value, 
and the fluctuations around such configuration do not contain any tachyon. 
This process describes the decay of the unstable D-brane system into a stable vacuum or a 
stable D-brane system. 

On the other hand, the stabilization of closed string tachyons is a harder problem to 
clarify. 
It would be because, in many cases, the closed string tachyons are not related to any 
unstable subsystem in the theory. 
Some interesting discussions have been done on this problem 
\cite{CostaGutperle}\cite{GutperleStrominger}\cite{RussoTseytlin}
\cite{Suyama}\cite{APS}\cite{Dabholkar}
, which seem to be right directions to understand the closed string tachyons. 

In some cases, the following background geometry plays an important role, 
\begin{equation}
ds^2 = ds^2_{flat} + dr^2 + r^2(d\theta+qdy)^2+dy^2.
  \label{KKMmetric}
\end{equation}
This is known as the Kaluza-Klein Melvin background \cite{Melvin}\cite{KKM}. 
Here $q$ is a real constant and the $y$-direction is compactified on $S^1$ with the radius 
$R$. 
This metric is locally flat, so this would be an exact background of string theory and 
M-theory. 
All of supersymmetries are broken in this background since the spin structure is changed by 
turning on $q$. 

The eleven-dimensional KK-Melvin can be interpreted as a Type IIA background, by the 
dimensional reduction along the $y$-direction. 
The Type IIA background has a localized R-R flux, known as a fluxbrane 
\cite{GutperleStrominger}, and 
the dilaton diverges far away from the core of the flux, so this background cannot be 
analyzed by perturbative calculations. 
It is conjectured in \cite{CostaGutperle} that Type IIA theory on this background is 
equivalent to Type 0A theory. 
The latter theory contains a closed string tachyon, and it is expected that Type 0A theory 
would decay into Type IIA theory on the flat spacetime, via a condensation of the closed 
string tachyon. 
Similar arguments are applied to non-supersymmetric heterotic theories \cite{Suyama}. 
There are also some investigations on closed string tachyons appeared in different contexts 
\cite{APS}\cite{Dabholkar}. 

A common property of the systems mentioned above is that they are superstring theories 
on supersymmetry-breaking backgrounds. 
Therefore it would be reasonable that such theories are stabilized to be the theories on 
supersymmetric vacua. 
Then one is led to think that the closed string tachyon condensation would deform the 
background geometry in a drastic way. 
So investigations of the mechanism of the tachyon condensation would give us many insights 
into the description of spacetime in string theory. 

\vspace{2mm}

In this paper, we investigate the properties of superstring theory on the ten-dimensional 
KK-Melvin in more detail. 
This system is interesting since closed string tachyons appear in a region of the parameter 
space of the background. 
In addition, the fate of them could be deduced; the system would decay into 
the same string theory on a supersymmetric background. 
Moreover, the system includes non-supersymmetric theories, i.e. Type 0 theory and the 
non-supersymmetric heterotic theories, as various limits of the system. 
For simplicity, we will mainly consider Type IIA theory on the KK-Melvin, but the results 
would be applicable to the heterotic cases. 
In fact, one interesting open question is the relation between the tachyon 
condensation and the gauge symmetry breaking in the heterotic theories \cite{Suyama}. 
So discussions for the heterotic cases are also relevant. 

The organization of this paper is as follows. 
We recall the worldsheet action for the KK-Melvin in section \ref{NLS}, and try to 
understand the pattern of the mass spectrum in the unwinding sector, in terms of 
the field theory language, in section \ref{unwind}. 
We also comment on marginal deformations of the worldsheet theory. 
Tachyons only appear in the winding sectors, so the similar discussions for the tachyons 
are difficult. 
We just specify the parameter region in which tachyons appear, in section \ref{wind}. 
In section \ref{LS}, we construct a linear sigma model of a string on the KK-Melvin. 
Then we investigate deformations of the background due to the tachyon 
condensation, by using the model, in section \ref{TCdeform}. 
This analysis has close similarity to \cite{APS}. 
In section \ref{het}, we briefly comment on the tachyon condensation and the gauge symmetry 
breaking in the heterotic theories.

\vspace{1cm}

\section{Nonlinear sigma model for the KK-Melvin} \label{NLS}

\vspace{5mm}

The action for Type IIA theory on the ten-dimensional KK-Melvin (\ref{KKMmetric}) 
is $S=S_{flat}+S_{Melvin}$ \cite{RT2}, where 
\begin{eqnarray}
S_{Melvin} &=& -\frac1{4\pi\alpha'}\int d^2\sigma\left\{
   (\partial_aX^\dagger-iq\partial_aYX^\dagger)(\partial_bX+iq\partial_bYX)
  +\partial_aY\partial_bY \right\}\eta^{ab} \nonumber \\
 & & +\frac i\pi\int d^2\sigma \left\{ 
   \psi_L^\dagger(\partial_-+iq\partial_-Y)\psi_L 
  +\psi_R^\dagger(\partial_++iq\partial_+Y)\psi_R \right. 
          \label{action} \\
 & & \hspace{2cm} \left.
  +\psi_L^Y\partial_-\psi_L^Y+\psi_R^Y\partial_+\psi_R^Y \right\}, \nonumber 
\end{eqnarray}
and $S_{flat}$ is the action for the free theory corresponding to the flat spacetime 
directions $x^\mu$ with $\mu=0, \cdots, 6$. 
We have employed the RNS formalism. 
The fields $X$, $\psi_{L,R}$ are defined as follows, 
\begin{eqnarray}
&& X = X^7+iX^8, \nonumber \\
&& \psi_{L,R} = \psi_{L,R}^7+i\psi_{L,R}^8. \nonumber 
\end{eqnarray}
Note that the vector indices of the fermions are the Lorentz indices, not the Einstein 
ones \cite{RT2}. 
The $y$-direction is compactified on $S^1$, so the field $Y$ is a compact boson, satisfying 
\begin{equation}
Y(\sigma+2\pi) = Y(\sigma) + 2\pi wR, 
\end{equation}
where $w$ is the winding number. 

The action (\ref{action}) can be reduced to a free action by the following field 
redefinitions, 
\begin{eqnarray}
&& X = e^{-iqY}\tilde{X}, \nonumber \\
&& \psi_{L,R} = e^{-iqY}\tilde{\psi}_{L,R}. \nonumber 
\end{eqnarray}
Then all fields satisfy the free wave equations. 
So the theory is described by the following free twisted fields, 
\begin{eqnarray}
&& \tilde{X}(\sigma+2\pi) = e^{2\pi iwqR}\tilde{X}(\sigma), \nonumber \\
&& \tilde{\psi}_{L,R}(\sigma+2\pi) = \pm e^{2\pi iwqR}\tilde{\psi}_{L,R}(\sigma),  \nonumber 
\end{eqnarray}
and free untwisted fields corresponding to the flat spacetime directions and the 
$y$-direction. 

The only difference from just a free theory is a zeromode $p$ of the field $Y$. 
Its canonical momentum $P_Y$ is 
\begin{eqnarray}
P_Y &\equiv& \frac{\delta S}{\delta \partial_\tau Y} \nonumber \\
    &=& \frac1{2\pi\alpha'}\partial_\tau Y
       +\frac{iq}{4\pi\alpha'}\left\{X(\partial_\tau X^\dagger-iq\partial_\tau YX^\dagger)
                                    -X^\dagger(\partial_\tau X+iq\partial_\tau YX)\right\}
       \nonumber \\
    & & -\frac q{2\pi}(\psi_L^\dagger\psi_L+\psi_R^\dagger\psi_R). 
\end{eqnarray}
It is the zeromode of $P_Y$ that is quantized, so the zeromode $p$ is 
\begin{eqnarray}
p &\equiv& \int_0^{2\pi}d\sigma \frac1{2\pi\alpha'}\partial_\tau Y \nonumber \\
  &=& \frac mR -qJ, 
\end{eqnarray}
where $m$ is an integer, and $J$ is the angular momentum operator in the 7-8 plane, 
\begin{eqnarray}
J &=& \int_0^{2\pi}d\sigma \left[
     \frac{i}{4\pi\alpha'}\left\{X(\partial_\tau X^\dagger-iq\partial_\tau YX^\dagger)
                                    -X^\dagger(\partial_\tau X+iq\partial_\tau YX)\right\}
     \right. \nonumber \\
  & & \left.
    -\frac 1{2\pi}(\psi_L^\dagger\psi_L+\psi_R^\dagger\psi_R) \right]. 
\end{eqnarray}

\vspace{1cm}

\section{Unwinding sector}  \label{unwind}

\vspace{5mm}

\subsection{Mass spectrum}

\vspace{5mm}

Consider first the unwinding sector. 
In this sector, $\tilde{X}$ and $\tilde{\psi}_{L,R}$ have the ordinary periodicities. 
So the effect of the nontrivial background appears only in the zeromode $p$. 
The mass operator and the level-matching condition are as follows, 
\begin{eqnarray}
\alpha'M^2 &=& \alpha'\left(\frac mR-qJ\right)^2+2(N+\tilde{N}-a), 
    \label{UTmass} \\
N &=& \tilde{N}, 
\end{eqnarray}
where $N$, $\tilde{N}$ are the ordinary level operators for the left- and the right-movers, 
respectively, and $a$ is a normal ordering constant. 

One can easily see, from the expression (\ref{UTmass}), that there is no tachyonic mode in 
the unwinding sector. 
In fact, as will be shown in section \ref{wind}, tachyons appear in the winding sectors. 
The mass of unwinding states are shifted according to their eigenvalues of $J$. 

Let's focus on some states in the NS-NS sector, which are massless when $q=0$. 
They are the graviton $G_{MN}$, the antisymmetric tensor $B_{MN}$ and the dilaton $\Phi$. 
The followings are the eigenvalues of $J$ for the states in the NS-NS sector, corresponding 
to the components of the above fields, 
\begin{equation}
\begin{array}{rl}
J=2: & G_{zz},\ B_{zz}, \\
1: & G_{\mu z},\ G_{yz},\ B_{\mu z},\ B_{yz}, \\
0: & G_{\mu\nu},\ G_{\mu y},\ G_{yy}, G_{z\bar{z}}, \\
   & B_{\mu\nu},\ B_{\mu y},\ \Phi, \\
-1:& G_{\mu\bar{z}},\ G_{y\bar{z}},\ B_{\mu\bar{z}},\ B_{y\bar{z}}, \\
-2:& G_{\bar{z}\bar{z}},\ B_{\bar{z}\bar{z}},
\end{array} 
\end{equation}
where $G_{\mu z}=(G_{\mu 7}-iG_{\mu 8})/2$, etc. 
For a small value of $q$, massless fields in the unwinding sector are the components of 
the above fields whose eigenvalues of $J$ are zero, and the other components become massive 
with the masses of order $q$. 

\vspace{3mm}

It is expected that properties of the unwinding states can be understood in terms of the 
field theory language. 
Since the above results are obtained from the tree level analysis, the mass shift 
discussed above will be seen by investigating the corresponding classical supergravity 
lagrangian. 
In fact, one needs only kinetic terms of the spacetime fields to see the mass shift. 

Consider a vector field $V^A$, for simplicity. 
Here $A$ is the Lorentz index. 
Since the massless fields correspond to the states created by the fermionic oscillators, 
their indices should be the Lorentz ones, as noted in the previous section. 
The kinetic term of $V^A$ is 
\begin{equation}
G^{MN}\nabla_MV^A\nabla_NV_A 
 = G^{MN}(\partial_MV^A+{{\omega_M}^A}_CV^C)(\partial_NV_A+{\omega_N}_{,AD}V^D), 
    \label{kinetic}
\end{equation}
where ${{\omega_M}^A}_B$ is the spin connection. 
For the metric (\ref{KKMmetric}), most of the components of the spin connection vanish, 
except for the followings, 
\begin{equation}
{\omega_y}^{78} = -{\omega_y}^{87} = -q. 
\end{equation}
Then $\nabla_{\mu,7,8}$ are reduced to the ordinary partial derivatives. 
The kinetic term (\ref{kinetic}) becomes 
\begin{equation}
\eta^{\mu\nu}\partial_\mu V^A\partial_\nu V_B+(\partial_7V^A)^2+(\partial_8V^A)^2
+\left(\nabla_yV^A-q(x^7\partial_8-x^8\partial_7)V^A\right)^2. 
\end{equation}
The last term can be rewritten as 
\begin{equation}
(\partial_yV^\mu-iqLV^\mu)^2+|\partial_yV^z+iqV^z-iqLV^z|^2+(\partial_yV^y-iqLV^y)^2, 
  \label{massterm}
\end{equation}
where $L=-i(x^7\partial_8-x^8\partial_7)$ is the orbital angular momentum, and 
$V^z = V^7+iV^8$. 
If the vector field $V^A$ is a simultaneous eigenstate of $\partial_y$ and $L$, 
then the terms (\ref{massterm}) become mass terms with the masses 
\begin{equation}
m^2 = \left\{
\begin{array}{ll}
\left(\frac kR-q(l-1)\right)^2 & \mbox{for $V^z$,} \\
\left(\frac {k'}R-ql\right)^2  & \mbox{for $V^\mu,\ V^y$}. 
\end{array}
\right.
\end{equation}
where $l$ is an eigenvalue of $L$. 
The corresponding result for an arbitrary bosonic field can be obtained in the same way. 
For example, a field $\phi$ with $n_z$ $z$-indices and $n_{\bar{z}}$ $\bar{z}$-indices has a 
mass 
\begin{equation}
m^2 = \left(\frac kR-q(l-(n_z-n_{\bar{z}}))\right)^2. 
  \label{FTmass}
\end{equation}
This value of $m^2$ coincides with the value of $p^2$ in the string spectrum. 
Note that the total angular momentum $j$ for the field $\phi$ should be 
$l+(n_z-n_{\bar{z}})$. 
This is not inconsistent with the result (\ref{FTmass}). 
The operator (\ref{UTmass}) acts on the states, and a state with $z$-index corresponds to 
a field with $\bar{z}$-index. 

The situation is the same for spinor fields. 
Spinor states have half-integral eigenvalues of $J$, so there is no massless 
fermion unless $qR$ is an even integer. 
In particular, at $qR=1$ all spinor states have masses of order at least $1/R$, while some 
massless bosons appear and the translational invariance is restored. 
So all spacetime fermions are infinitely massive in the $R\to 0$ limit, 
which corresponds to the Type 0A limit \cite{CostaGutperle}\cite{Type0}. 

\vspace{3mm}

As mentioned before, the KK-Melvin breaks all of supersymmetries. 
The restoration of supersymmetry at $q=0$ can be easily understood; some of massive fermions 
(and also bosons) just become massless, and degrees of freedom of bosons and fermions 
become the same. 
Since Type 0A theory can be interpreted as Type IIA theory on the KK-Melvin with $qR=1$, the 
restoration of supersymmetry in this theory after the tachyon condensation would not be a 
difficult problem, in view of the dual Type IIA picture; 
infinitely massive fermions become massless. 
This would suggest that, in Type 0A theory, some nonperturbative fermionic states become 
massless after the tachyon condensation. 
It is very interesting to clarify the mechanism for the restoration of supersymmetry in view 
of Type 0A theory.

\vspace{5mm}

\subsection{Marginal deformations of the worldsheet theory}  \label{marginal}

\vspace{5mm}

In the CFT viewpoint, the tachyon condensation is described by a relevant deformation. 
As discussed in \cite{CostaGutperle}\cite{KKM}, a theory on the KK-Melvin would decay into 
the theory on the flat spacetime. 
So, naively, the tachyon condensation seems to be described by a marginal deformation 
which modifies the background field configurations. 
In this section, we discuss some marginal deformations of Type IIA theory on the KK-Melvin. 

The condensation of a massless state corresponds to a marginal deformation of the theory. 
We focus on the condensation of the gravitons. 
The massless components of the metric for generic $q$ are 
\begin{equation}
G_{\mu\nu},\ G_{\mu y},\ G_{yy},\ G_{z\bar{z}}.
\end{equation}
The last two would be important for the deformation of the KK-Melvin. 

The condensation of the state corresponding to $G_{yy}$ changes the coefficient of 
$(\partial Y)^2$ in $S_{Melvin}$. 
Denote the coefficient as $C$. 
Then, by rescaling $Y$ as 
\begin{equation}
Y = \frac1{\sqrt{C}}Y', 
\end{equation}
the action becomes again the same form as $S_{Melvin}$, with different parameters $q'$ and 
$R'$, 
\begin{equation}
q'=\frac q{\sqrt{C}}, \hspace{1cm} R'= \sqrt{C}R. 
\end{equation}
Therefore, this deformation changes the radius $R$ while keeping $qR$ fixed. 
This seems natural since the combination $qR$ is related to the spin structure of the 
background geometry. 

The deformation corresponding to $G_{z\bar{z}}$ is in fact equivalent to that of $G_{yy}$. 
Remember that the indices of the spacetime fields are the Lorentz indices. 
So the condensation would change the coefficient of $|\partial\tilde{X}|^2$, not 
$|\partial X|^2$, in $S_{Melvin}$. 
Thus this is equivalent to the rescaling of $Y$. 

Interestingly, as shown above, marginal deformations of the KK-Melvin change the parameters 
in a restricted manner. 
In particular, changing only $q$ is not marginal. 
This seems consistent with the fact that pair creations of D-branes, which is certainly not 
an on-shell phenomenon, decrease the value of $q$ \cite{CostaGutperle}\cite{KKM}. 
In section \ref{TCdeform}, we will discuss the tachyon condensation on the KK-Melvin. 
It will be shown that the phenomenon investigated there does not correspond to the marginal 
deformations of the worldsheet theory discussed above.

\vspace{1cm}

\section{Tachyons in the winding sectors}  \label{wind}

\vspace{5mm}

Tachyonic modes appear in the winding sectors. 
They are only in the NS-NS sector \cite{RT2}. 
Since the boundary condition of $\tilde{X}$ is twisted, the tachyons do not have zeromodes 
along the 7,8-directions. 
This can be understood in terms of the geometry of the KK-Melvin. 
One can see that the $S^1$ in the KK-Melvin is twisted, that is, the translation along the 
$S^1$ is accompanied by a rotation in the 7-8 plane. 
So winding strings are also winding around the origin of the 7-8 plane. 
Thus the winding states are localized in the 7-8 plane. 
In general, properties of winding states are stringy, so it would be difficult to 
understand them in terms of the field theory language. 
One way to analyze them is to see the T-dual theory, in which tachyons appear in the 
unwinding sector \cite{Tdual}. 
Such arguments for the heterotic theories have been done in \cite{Suyama}. 

In this section, we will just specify the parameter region in which tachyons appear. 
The perturbative spectrum of this system is already analyzed in \cite{RT2}. 

It can be shown that the investigations for the $w=1$ case are sufficient. 
The tachyonic region in the parameter space for any $w\ne 1$ sector is included in the 
region for $w=1$. 

The mass operator and the level-matching condition in the $w=1$ sector are as follows, 
\begin{eqnarray}
&& \alpha'M^2 = \alpha'p^2+\frac{(wR)^2}{\alpha'}+2(N_{free}+\tilde{N}_{free}
   +N_{qR}+\tilde{N}_{qR}+N_Y+\tilde{N}_Y+a), \\
&& N_{free}+N_{qR}+N_Y-\tilde{N}_{free}-\tilde{N}_{qR}-\tilde{N}_Y
    = -m+RJ, 
\end{eqnarray}
where $N_{free}$, $N_Y$ etc. are the ordinary level operators for the fields corresponding 
to the $x^\mu$- and $y$-directions, respectively. 
The level operators $N_{qR}$ and $\tilde{N}_{qR}$ for the twisted fields have the following 
forms, 
\begin{eqnarray}
&& N_{qR} = \sum_{k\in {\bf Z}}:\alpha_{n-qR}^\dagger\alpha_{n-qR}:
           +\sum_{k\in {\bf Z}}\left(n+\frac12-qR\right)
                               :b_{n+\frac12-qR}^\dagger b_{n+\frac12-qR}:,
     \nonumber \\
&& \tilde{N}_{qR} = \sum_{k\in {\bf Z}}:\alpha_{n+qR}^\dagger\alpha_{n+qR}:
           +\sum_{k\in {\bf Z}}\left(n+\frac12+qR\right)
                               :b_{n+\frac12+qR}^\dagger b_{n+\frac12+qR}:,
     \nonumber 
\end{eqnarray}
where the mode operators satisfy 
\begin{equation}
\begin{array}{ll}
[\alpha_{m-qR}, \alpha_{n-qR}^\dagger] = (m-qR)\delta_{mn}, & 
[\alpha_{m+qR}, \alpha_{n+qR}^\dagger] = (m+qR)\delta_{mn}, \\
\{b_{m-qR}, b_{n-qR}^\dagger\} = \delta_{mn}, & 
\{b_{m+qR}, b_{n+qR}^\dagger\} = \delta_{mn}, 
\end{array}
\end{equation}
and $a$ is a normal ordering constant. 

The mass spectrum is periodic in $qR$ with the period two. 
So we restrict the parameter region to $0\le qR \le 2$. 
In this region, the normal ordering constant $a$ takes the following values, 
\begin{equation}
a=\left\{
\begin{array}{ll}
qR-1, & \left(0<qR<\frac12\right) \\
-qR,  & \left(\frac12<qR<1\right) \\
qR-2, & \left(1<qR<\frac32\right) \\
-qR+1.& \left(\frac32<qR<2\right)
\end{array}
\right.
\end{equation}
Note that this is the only negative contribution to the mass squared. 
Since the constant $a$ is not less than $-1$, a state is not tachyonic if the state is 
created by oscillators of the fields except for $\tilde{X}$, $\tilde{\psi}_{L,R}$. 

\begin{figure}[htb]
 \epsfxsize=20em
 \centerline{\epsfbox{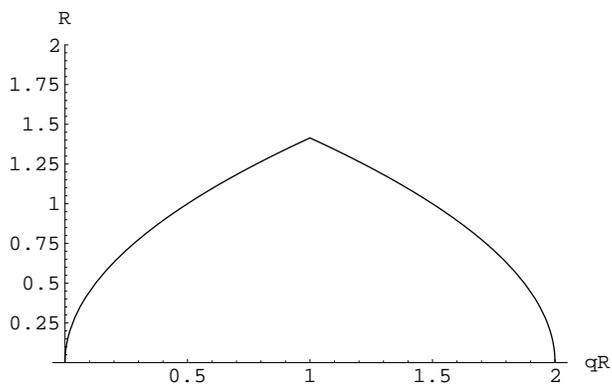}}
 \caption{Tachyonic region}
   \label{fig}
\end{figure}

The lightest state in the spectrum is 
\begin{eqnarray}
&& b_{\frac12-qR}^\dagger b_{qR-\frac12}|m=0\rangle_{NS\mbox{-}NS}, 
   \hspace{1cm} \left(0<qR<\frac12\right) \nonumber \\
&& b_{\frac32-qR}b_{qR-\frac32}^\dagger|m=0\rangle_{NS\mbox{-}NS}, 
   \hspace{1cm} \left(\frac32<qR<2\right) \nonumber \\
\end{eqnarray}
and $|m=0\rangle_{NS\mbox{-}NS}$ for the other region, in which the GSO-projection is 
reversed. 
If the state has a non-negative mass squared, then the theory is tachyon-free. 
The line on which the lightest state becomes massless is determined as follows, 
\begin{equation}
\left\{
\begin{array}{ll}
R^2 = 2\alpha'qR, & (0<qR<1) \\
R^2 = 2\alpha'(2-qR). & (1<qR<2)
\end{array}
\right.
   \label{line}
\end{equation}
This is shown in figure \ref{fig}. 
When the parameters of the KK-Melvin correspond to a point under this line, there exist 
tachyonic modes in the winding sectors. 
Note that the point $(R,qR)=(0,1)$ corresponds to the Type 0A limit. 

\vspace{3mm}

Any point in figure \ref{fig} represents a string theory on a KK-Melvin. 
If one chooses a point in the tachyonic region, the corresponding theory is unstable, and it 
would decay into some stable theory without any tachyon. 
It is conjectured in \cite{CostaGutperle}\cite{GutperleStrominger} that the endpoint of the 
decay would be a supersymmetric vacuum which corresponds to a point on the line $qR=0$ or 
equivalently $qR=2$. 
In the next section, we will discuss which point is the endpoint of the decay, in more 
detail. 
In particular, we will focus on the change of $R$ through the decay. 

\vspace{1cm}

\section{Linear sigma model}  \label{LS}

\vspace{5mm}

In this section, we would like to discuss the closed string tachyon condensation in string 
theory on the KK-Melvin. 
An interesting approach is investigated in \cite{APS} toward the understanding of a similar 
phenomenon. 
They analyze, by using a D-brane probe, Type II theory compactified on non-compact orbifolds 
in which tachyons appear in the twisted sectors. 
And they conjecture that the condensation of the tachyons deforms the background geometry, 
and finally the background becomes a supersymmetric one. 
It seems natural that the tachyon condensation changes the background spacetime since both 
tachyons and gravitons are closed string states. 
Therefore, to analyze such phenomena, one should have to employ a formalism in which the 
dependence on the background geometry is implicit. 

We will use a linear sigma model \cite{LSM} to investigate string theory on the KK-Melvin. 
The model is not related to any D-brane, so the result we will obtain below would be 
applicable to the heterotic theories. 
The structure of the model is very similar to the worldvolume theory of the probe D-brane 
discussed in \cite{APS}, so the similar computations can be applied to our model.

\vspace{5mm}

\subsection{KK-Melvin as a quotient space}

\vspace{5mm}

It is convenient for our purpose to describe the KK-Melvin in a slightly different way. 
We consider the case with $qR=k+1/n$, where $k$ is an integer and $n$ is a positive integer. 
Then the corresponding KK-Melvin can be obtained as a ${\bf Z}_n$ quotient space. 

Recall that the KK-Melvin can be obtained from the flat spacetime, 
\begin{equation}
ds^2 = \eta_{\mu\nu}dx^\mu dx^\nu+dr^2+r^2d\tilde{\theta}+dy^2, 
\end{equation}
by the following identifications, 
\begin{eqnarray}
&& y \sim y+2\pi mR, 
      \label{identify1} \\
&& \tilde{\theta} \sim \tilde{\theta}+2\pi l+2\pi m\frac1n, 
      \label{identify2}
\end{eqnarray}
where $m,l$ are integers. 

Every integer $m$ can be written uniquely as 
\begin{equation}
m = rn+s, \hspace{1cm} (r,s\in {\bf Z},\ 0\le s<n).
\end{equation}
So the identifications (\ref{identify1})(\ref{identify2}) are rewritten as follows, 
\begin{eqnarray}
&& y \sim y+2\pi rnR+2\pi sR, \nonumber \\
&& \tilde{\theta} \sim \tilde{\theta}+2\pi l+2\pi r+2\pi\frac sn. \nonumber 
\end{eqnarray}
When $s=0$, these identifications are nothing but the $S^1$ compactification with the radius 
$nR$. 
Therefore the KK-Melvin can be expressed as 
\begin{equation}
{\bf R}^7\times \left( {\bf R}^2\times S^1/{\bf Z}_n \right). 
\end{equation}
The action of an element $\sigma_s\in {\bf Z}_n$ is 
\begin{equation}
\sigma_s : \left\{
\begin{array}{l}
y \mapsto y+2\pi sR, \\ \tilde{\theta} \mapsto \tilde{\theta}+2\pi s/n.
\end{array}
\right.
   \label{Zn}
\end{equation}
This expression is useful to construct a linear sigma model of the KK-Melvin, which will be 
shown below.

\vspace{5mm}

\subsection{A model}

\vspace{5mm}

We describe a linear sigma model of the KK-Melvin in this subsection. 
The model for the flat spacetime directions is trivial, so we only consider the 
three-dimensional part corresponding to the worldsheet action $S_{Melvin}$. 
We will restrict ourselves to considering bosonic degrees of freedom in the KK-Melvin. 

The linear sigma model we analyze below is a two-dimensional gauge theory. 
The gauge group is $U(1)^{n+1}$, whose gauge fields are $A^i$ $(i=0,\cdots,n-1)$ and $A$. 
Note that one linear combination of $A^i$ decouples. 
There are also $n+2$ scalar fields $z_0,\cdots,\ z_{n-1}$ and $w_1,\ w_2$. 
They are charged under the gauge fields. 
The scalar $z_i$ has the charge $+1$ for $A^i$ and $-1$ for $A^{i+1}$ (the indices have to 
be considered modulo $n$), and the scalars $w_1,\ w_2$ both have the charge $+1$ for $A$. 
The potential is given as follows, 
\begin{eqnarray}
V &=& \sum_{i=0}^{n-2}\left(|z_i|^2-|z_{i+1}|^2\right)^2+\left(|z_{n-1}|^2-|z_0|^2\right)^2
        \nonumber \\
  & & +\left(|w_1|^2-\zeta^2\right)^2+\left(|w_2|^2-\zeta^2\right)^2, 
    \label{potential} 
\end{eqnarray}
where $\zeta>0$. 

In the IR limit, this theory flows into a nonlinear sigma model whose target space is the 
moduli space ${\cal M}$ of the theory. 
The moduli space is the space of vevs of the scalar fields which minimize the potential, 
moded out by the gauge symmetry. 
The potential has minima at 
\begin{eqnarray}
&& |z_0| = \cdots = |z_{n-1}|, \nonumber \\
&& |w_1| = |w_2| = \zeta. \nonumber 
\end{eqnarray}
We fix the gauge symmetry by the following conditions, 
\begin{equation}
\arg(z_0) = \cdots = \arg(z_{n-1}) = \arg(w_1). 
\end{equation}
Thus the vevs of the scalar fields are now written as 
\begin{eqnarray}
&& z_0 = \cdots = z_{n-1} \equiv z \in {\bf C}, \nonumber \\
&& w_1 = \zeta\frac z{|z|}, \hspace{5mm} w^2 = \zeta e^{i\phi}. \nonumber 
\end{eqnarray}
So there is a surjection from the space ${\bf C}\times S^1$ onto the moduli space 
${\cal M}$. 

There is a residual ${\bf Z}_n$ symmetry whose generator acts on the scalar fields as 
\begin{equation}
z_i \to e^{\frac{2\pi i}n}z_i, \hspace{5mm} w_{1,2} \to e^{\frac{2\pi i}n}w_{1,2}. 
\end{equation}
They can be rewritten as follows, 
\begin{equation}
\arg(z_i) \to \arg(z_i)+2\pi\frac1n, \hspace{5mm} \phi \to \phi+2\pi\frac1n. 
   \label{twistID}
\end{equation}
The action of elements of ${\bf Z}_n$ coincides with (\ref{Zn}), up to scale, so the moduli 
space 
\begin{equation}
{\cal M} = {\bf C}\times S^1/{\bf Z}_n
\end{equation}
is the KK-Melvin with $qR=1/n$ modulo integer. 

\vspace{3mm}

The sigma model metric can also be calculated. 
We parametrize the potential minima as 
\begin{equation}
z_i = re^{i\theta_i}, \hspace{5mm} w_1 = \zeta e^{i\phi_1}, \hspace{5mm} 
w_2 = \zeta e^{i(\phi_2+\frac1n\theta)}, 
\end{equation}
where $\theta=\sum_{i=0}^{n-1}\theta_i$. 
The phase of $w_2$ is shifted so as to take into account the twisted identifications 
(\ref{twistID}). 
Then the kinetic terms for the scalars become 
\begin{eqnarray}
&& \sum_{i=0}^{n-2}\left\{ (\partial r)^2+r^2(\partial \theta_i-(A^i-A^{i+1}))^2\right\}
  +\left\{(\partial r)^2+r^2(\partial\theta_{n-1}-(A^{n-1}-A^0))^2\right\}
     \nonumber \\
&& \hspace{1cm} +\zeta^2(\partial\phi_1-A)^2
                +\zeta^2\left(\partial\phi_2+\frac1n\partial\theta-A\right)^2.
     \nonumber 
\end{eqnarray}
The first two terms coincide with the kinetic terms in the worldvolume theory of the probe 
D-brane \cite{APS}, so the gauge fields $A^i$ can be integrated out in the same way. 
Another gauge field $A$ can also be integrated out easily, and the result is 
\begin{equation}
\frac{\zeta^2}2\left(\partial\phi+\frac1n\partial\theta\right)^2, \hspace{1cm} 
\phi=\phi_2-\phi_1. 
\end{equation}
Thus we obtain the following metric 
\begin{equation}
ds^2 = ndr^2+\frac{r^2}nd\theta^2+\frac{\zeta^2}2\left(d\phi+\frac1nd\theta\right)^2,
  \label{preKKM}
\end{equation}
from the resulting expression of the kinetic terms. 

\vspace{2mm}

Let us show the relation between the metric (\ref{preKKM}) and the KK-Melvin. 
The metric can be rewritten as 
\begin{equation}
ds^2 = n\left[dr^2+\frac{r^2}{n^2}d\theta^2
      +\left(\frac\zeta{\sqrt{2n}}d\phi+\frac\zeta{\sqrt{2n^3}}d\theta\right)^2\right]. 
\end{equation}
Let $\zeta\phi/\sqrt2=ny$. 
The factor $n$ is in front of $y$, since the radius of $S^1$ in the space 
${\bf C}\times S^1$, which corresponds to the $\phi$-direction, 
is $n$ times larger than that of the $y$-direction in the KK-Melvin, as mentioned in the 
previous subsection. 
From this identification, the radius of the $y$-direction is 
\begin{equation}
R = \frac\zeta{\sqrt2}n^{-\frac32}. 
\end{equation}
Finary the metric becomes 
\begin{equation}
ds^2 = n\left[ dr^2+\frac{r^2}{n^2}d\theta^2+(ndy+Rd\theta)^2 \right]. 
\end{equation}
This exactly coincides, up to an overall scale, with the nontrivial part of the metric of 
the KK-Melvin in a coordinate system shown in \cite{GutperleStrominger}, with $qR=1/n$ 
modulo integer. 
Therefore we conclude that the linear sigma model we have discussed so far flows into the 
nonlinear sigma model of the KK-Melvin in the IR limit. 
Note that the radius $R$ depends on $n$. 
This will be important later. 

Since we have considered only bosonic degrees of freedom, one cannot 
distinguish the KK-Melvin with $qR=1+1/n$ from that with $qR=1/n$, within this model. 
The difference would become apparent when fermionic degrees of freedom are included in the 
model.

\vspace{1cm}

\section{Deformations of the model}  \label{TCdeform}

\vspace{5mm}

Now we discuss the closed string tachyon condensation in the KK-Melvin by analyzing effects 
of deformations of the potential of the linear sigma model. 

Suppose that a tachyon vev with some profile is turned on. 
This would be described by a relevant deformation of the theory. 
Relevant deformations change the IR behavior of a theory. 
In a linear sigma model, the IR behavior is determined by the minima of the potential, 
so relevant deformations would correspond to perturbations of the potential which change 
the potential minima. 

Naively, forms of such relevant deformations can be deduced from the following simplified 
argument. 
Assume a potential $V(x)=(x^2-a^2)^2$. 
For simplicity, we assume that the system is symmetric under the sign flip of $x$; $x\to-x$. 
Then the generic perturbation of the potential is a polynomial $f(x^2)$. 
This polynomial can be rewritten as 
\begin{equation}
f(x^2) = (x^2-a^2)^2g(x^2)+Ax^2+B. 
\end{equation}
Thus this means that, in general, the only $x^2$ term is essential for the deformations of 
the potential minima. 
So this term would correspond to a relevant deformation of the theory. 

Similar situation can be seen in the Landau-Ginzburg model. 
In this model, it is shown that adding the lowest order term corresponds to the most 
relevant deformation of a CFT. 

Therefore it seems natural to expect that adding one of the lowest order terms in our linear 
sigma model would correspond to a relevant deformation which would describe turning on a 
tachyon vev. 

We consider the following terms, 
\begin{equation}
\sum_{i=0}^{n-1}m_i|z_i|^2+\mu_1|w_1|^2+\mu_2|w_2|^2, 
  \label{perturbation}
\end{equation}
with $\sum_{i=0}^{n-1}m_i=0$. 
A reason for imposing this condition on $m_i$ is as follows: 
The condition implies that we exclude the perturbation of the form $\sum|z_i|^2$. 
This perturbation lifts the moduli space and the dimension of the moduli space becomes 
smaller, even if the coefficient of the perturbation is small. 
We expect that the coefficients of the perturbations are related to the magnitude of the 
tachyon vev, and the effect of the tachyon condensation is continuous for the tachyon vev. 
So this perturbation would have nothing to do with the tachyon condensation. 

The last two terms in (\ref{perturbation}) just change the potential of $w_{1,2}$ in 
(\ref{potential}) into the form 
\begin{equation}
(|w_1|^2-\zeta_1^2)^2+(|w_2|^2-\zeta_2^2)^2+\mbox{const.}
\end{equation}
The metric obtained from this deformed potential is essentially the same as (\ref{preKKM}). 
The difference is the value of $\zeta$ which is now determined by the equation 
\begin{equation}
\frac1{\zeta^2} = \frac12\left(\frac1{\zeta_1^2}+\frac1{\zeta_2^2}\right). 
\end{equation}
Therefore this perturbation only changes the radius $R$ while keeping $qR$ fixed. 
So this would rather describe the marginal deformation corresponding to the change of 
$G_{yy}$, which is discussed in section \ref{marginal}. 

Let's focus on the other terms in (\ref{perturbation}). 
The effect of this perturbation to the metric is investigated in \cite{APS}, and we can use 
the result in our case. 
The resulting metric is 
\begin{equation}
ds^2 = n(r)dr^2+\frac{r^2}{n(r)}d\theta^2
      +\frac{\zeta^2}2\left(d\phi+\frac1nd\theta\right)^2. 
\end{equation}
We have used the same parametrization for $w_{1,2}$ as before. 
The function $n(r)$ behaves as 
\begin{equation}
n(r)\to \left\{
\begin{array}{ll}
n, & (r\to \infty) \\ n', & (r= 0)
\end{array}
\right.
\end{equation}
where $n'$ is an integer satisfying $1\le n'<n$. 
The integer $n'$ is determined from the values of $m_i$. 
The value of $qR$ at $r=0$ is thus different from the value at $r\to \infty$. 

This result also indicates that the radius of the $y$-direction varies as $r$ goes from 
zero to infinity, 
\begin{equation}
R\to \left\{
\begin{array}{ll}
\frac\zeta{\sqrt{2n^3}}, & (r\to\infty) \\ \frac\zeta{\sqrt{2n'n^2}}. & (r= 0)
\end{array}
\right.
\end{equation}
Therefore this shows that $R(r\to \infty)<R(r= 0)$. 

Recall that the closed string tachyons are localized around $r=0$. 
Thus the geometry around $r=0$ would be the one deformed by the tachyon condensation. 
So we could conclude that, after the tachyon condensation, the parameters $qR$, $R$ become 
larger. 
The system with $qR=2$ is the same theory as that with $qR=0$. 
Therefore string theory on the KK-Melvin would decay into the theory on the flat spacetime 
times $S^1$, and supersymmetry would be restored. 
When the tachyon condensation is fine-tuned so that the endpoint of the decay is a 
tachyon-free KK-Melvin, the theory then continues to decay via an instanton effect 
\cite{KKM}. 

\vspace{2mm}

The details of the above results we have obtained so far might not be correct. 
However, we expect that the qualitative picture is still meaningful. 
It is based on the following arguments: 
The ten-dimensional KK-Melvin can be lifted to the eleven-dimensional one, and it is related 
in \cite{GutperleStrominger} to a flat cone background in ten dimensions with a constant R-R 
field. 
If the arguments in \cite{APS} can be applied to this case, although a R-R background 
exists, one could claim that the scale of the radial direction is reduced, in addition to 
the change of the value of $qR$. 
So the relative scale of the $y$-direction would become larger, that is, the way of changing 
the parameters would match what we have found from the linear sigma model. 

\vspace{2mm}

There are some pieces of evidence which support our results. 

\vspace{1mm}

\hspace*{-7mm}(i) The tachyon condensation tends to drive the theory toward a tachyon-free 
theory, which can be seen from the figure \ref{fig}.  

\vspace{1mm}

\hspace{-7mm}(ii) It is conjectured in \cite{Type0} that Type 0A theory is decompactified 
to be M-theory after the tachyon condensation. 
This is based on an argument on the nonzero cosmological constant in the theory. 
Our result is consistent with the conjecture. 

\vspace{1mm}

\hspace{-7mm}(iii) As pointed out in \cite{Suyama}, the closed string tachyon condensation 
should induce the gauge symmetry breaking in heterotic theories. 
Consider a non-supersymmetric heterotic theory. 
This can be interpreted as a supersymmetric heterotic theory on the KK-Melvin with a Wilson 
line \cite{Suyama}, unless the gauge group is $E_8$. 
If the radius of $S^1$ of the KK-Melvin becomes large while keeping the Wilson line fixed, 
then the gauge symmetry is broken by the Wilson line. 
Note that the tachyon condensation would also reduce the rank of the gauge group, not just 
break the gauge group to a subgroup with the same rank. 
We will comment on this point in the next section.

\vspace{1cm}

\section{Tachyon condensation in heterotic theories}  \label{het}

\vspace{5mm}

As has been pointed out, when closed string tachyons condense in heterotic theories, the 
gauge symmetry should be broken in general, since the tachyons couple to the gauge fields 
in a nontrivial way. 
Moreover, it is expected that the rank of the gauge group is reduced by the tachyon 
condensation. 
However, it is well-known that consistency conditions of the theory severely restrict the 
rank of the gauge group. 
Is it really possible for non-supersymmetric heterotic theories to decay into supersymmetric 
heterotic theories ?

Recall the spectra of the non-supersymmetric theories. 
There are seven such theories characterized by their gauge groups; the $SO(16)\times SO(16)$ 
theory is tachyon-free, the theories with $SO(32),\ E_8\times SO(16),\ SO(8)\times SO(24),\ 
(E_7\times SU(2))^2,\ U(16),\ E_8$ all have tachyons. 
First four theories have tachyons in nontrivial representations of the gauge groups. 

Consider the $SO(32)$ theory. 
The tachyons form the vector representation of the $SO(32)$. 
Suppose that the tachyons acquire generic vevs. 
By a rigid gauge transformation, one can put them into the form, 
\begin{equation}
T = \left(
\begin{array}{c}
* \\ 0 \\ \vdots \\ 0
\end{array}
\right).
\end{equation}
So the gauge group is broken from $SO(32)$ to $SO(31)$. 
Therefore the rank of the gauge group is reduced by one, due to the tachyon condensation. 
The similar reduction would occur in other theories. 
The reduction of the rank is at most one. 

From the above considerations, one may think that the $SO(32)$ theory would decay into the 
supersymmetric heterotic theory compactified on $S^1$, whose gauge group is $SO(31)$, or its 
subgroup with the rank fifteen. 
However, such heterotic theory does not known. 
The rank of the gauge group less than sixteen is at most eight \cite{CHL}. 
So it may be concluded that the above non-supersymmetric heterotic theory would not be 
well-defined, since its stable vacuum seems not to exist. 

Similar situations are found in the supersymmetric heterotic theories on the KK-Melvin, even 
with a finite radius of the $S^1$. 
And in this case, there would be a possible solution of this puzzle. 
Recall that the rank of the gauge group can be larger than sixteen at special points in the 
moduli space of compactified heterotic theories \cite{enhance}. 
For example, the supersymmetric $SO(32)$ theory on $S^1$ can have the gauge group $SO(34)$ 
by choosing appropriate radius and Wilson line. 
Therefore it would be possible that the rank is sixteen even after the rank is reduced by 
one due to the tachyon condensation. 
The mechanism of this phenomenon, if exists, would be very complicated and the field 
theoretic interpretation would be difficult. 

Of course there are other possibilities. 
One is that the endpoint of the decay is a CHL string \cite{CHL}, by further reducing the 
rank in some way. 
And one cannot exclude the possibility that the theory becomes a non-critical string. 
This might be preferred because of the c-theorem in CFT. 

We expect that the first possibility would be realized. 
We think it natural because the theories we have considered are supersymmetric theories on a 
particular background. 
So the stable vacuum of such theories would probably be the supersymmetric vacuum.

\vspace{1cm}

{\Large {\bf Acknowledgements}}

\vspace{5mm}

I would like to thank S. Iso, S. Kawamoto, T. Lee, T. Matsuo, K. Okuyama, Y.Shibusa and  P. Yi 
for valuable discussions.

\end{document}